\documentclass{aa}
\usepackage{graphicx}
\usepackage{txfonts}

\begin{document}

   \title{A double radio halo in the close pair of galaxy clusters Abell 399 and Abell 401}

 \author{M. Murgia\inst{1,2}
          \and
          F. Govoni\inst{1}
          \and
          L. Feretti\inst{2}
         \and
          G. Giovannini\inst{2,3}
          }
   \institute{
              INAF - Osservatorio Astronomico di Cagliari, Poggio dei Pini, Strada 54, I--09012 Capoterra (CA), Italy
           \and
              INAF - Istituto di Radioastronomia, Via Gobetti 101, I--40129 Bologna, Italy
           \and
              Dipartimento di Astronomia, Univ. Bologna, Via Ranzani 1, I--40127 Bologna, Italy   
}

   \date{Received; accepted}

 
  \abstract
  {} 
   { Radio halos are faint radio sources usually located at the center of merging clusters of galaxies. These diffuse radio sources are rare, having so far been found only in about 30 clusters of galaxies, suggesting that particular conditions are needed to form and maintain them. It is interesting to investigate the presence of radio halos in close pairs of interacting clusters in order to possibly clarify their origin in relation to the evolutionary state of the merger. In this work, we study the case of the close pair of galaxy clusters A399 and A401. 
}
   {A401 is already known to contain a faint radio halo, while a hint of diffuse emission in A399 has been suggested based on the NVSS. To confirm this possibility, we analyzed deeper Very Large Array observations at 1.4 GHz of this cluster.}
   {We find that the central region of A399 is permeated by a diffuse low-surface brightness radio emission that we classify as a radio halo with a linear size of about 570 kpc and a central brightness of 0.3 $\mu$Jy/arcsec$^2$. Indeed, given their comparatively small projected distance of $\sim 3$ Mpc, the pair of galaxy clusters A401 and A399 can be considered as the first example of double radio halo system. The discovery of this double halo is extraordinary given the rarity of these radio sources in general and given that current X-ray data seem to suggest that the two clusters are still in a pre-merger state. Therefore, the origin of the double radio halo is likely to be attributed to 
the individual merging histories of each cluster separately, rather than to the result of a close encounter between the two systems.}
   {}


\titlerunning{A double radio halo in the close pair of galaxy clusters A399 and A401}

   \maketitle

\section{Introduction}

An ever increasing number of galaxy clusters exhibit
Mpc-scale synchrotron radio halos associated with the
intracluster medium. These elusive radio sources are located at the cluster center and are characterized by a
regular shape and extremely low surface brightness at levels of $\sim 1 \mu$Jy/arcsec$^2$ 
at 1.4 GHz. To date, about 30 radio halos are known (see Giovannini et al. 2009 for recent
compilation), and most have been found in clusters that show significant
evidence of an ongoing merger (e.g. Buote 2001).
The rarity of radio halos seems to suggest that particular conditions are required for their formation 
and maintenance. Merger events may play a
 strong role in the re-acceleration of the radio-emitting relativistic particles, 
thus providing the energy that powers these extended sources (e.g. Brunetti et al. 2009). 
In this context, it is interesting to investigate radio halos in close pairs of interacting clusters of galaxies. These systems may share the same recent dynamical history, so they are unique laboratories for studying the formation of radio halos and possibly clarifying the origin of these radio sources
 in relation to the evolutionary state of the merger.
\begin{figure*}[t]
\centering
\includegraphics[width=18 cm]{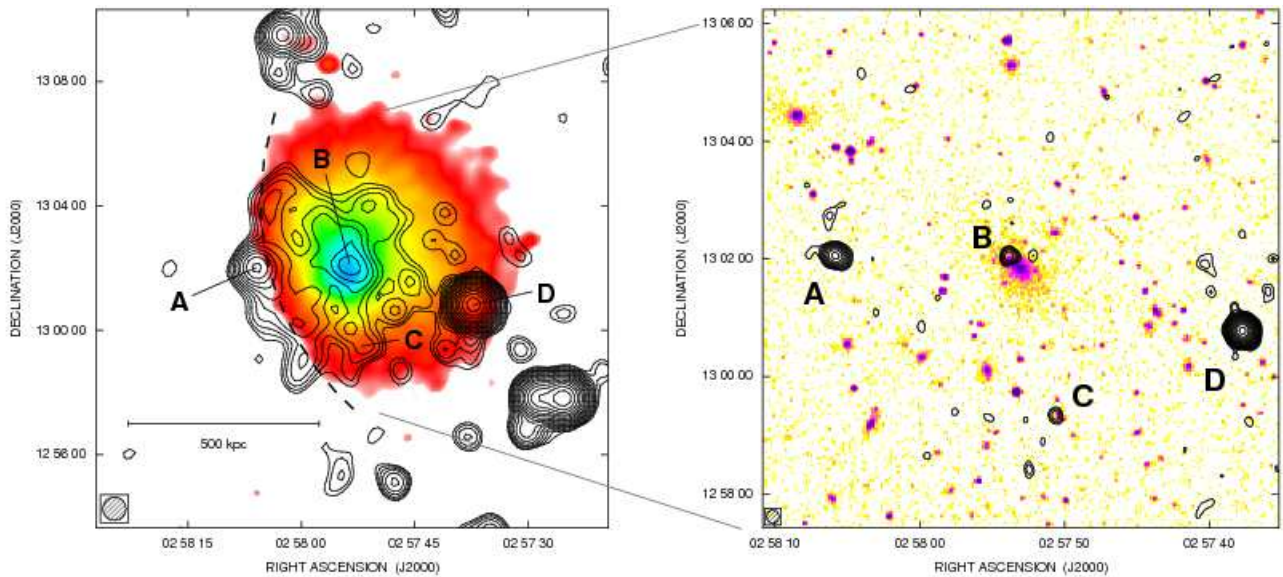}
\caption{Left: total intensity radio contours of A399 at 1.4 GHz
with the VLA in D configuration.
The image has an FWHM of 45$''\times 45''$.
The first contour level is drawn at 120 $\mu$Jy/beam
and the rest are spaced by a factor of $\sqrt{2}$.
The sensitivity (1$\sigma$) is 40 $\mu$Jy/beam.
The total intensity radio contours overlaid on the XMM X-ray image in the 0.2$-$12 keV band. The X-ray image has been convolved with a Gaussian of $\sigma=12\arcsec$. 
Right: zoom of the total intensity radio contours in the center of A399 
at 1.4 GHz with the VLA in C array.
The image has an FWHM of 14$''\times 13''$. 
The first contour level is drawn at 120 $\mu$Jy/beam
and the others are spaced by a factor of $\sqrt{2}$. The sensitivity 
(1$\sigma$) is 40 $\mu$Jy/beam. The contours of the radio intensity are 
overlaid on the optical DSS2 image.}
\label{fig1}
\end{figure*}

In this work we analyze the close pair of galaxy clusters A399 and A401, which
are located at a redshift of $z=0.071806$ and $z=0.073664$ (Oegerle \& Hill 2001).
This is an exceptional pair, since the two clusters are separated in projection by an angular distance of 36\arcmin, corresponding to a linear 
separation of 3 Mpc in our chosen cosmology (see below). A399 and A401 are two rich clusters of galaxies
with a similar average gas temperature of $kT\simeq 7$ and $kT\simeq 8$ keV, respectively.
The X-ray luminosities in the 0.1$-$2.4 keV band for the two clusters are $L_{X}=3.8$ and $6.5 \times10^{44}$ erg/s 
(Reiprich \& B{\"o}hringer 2002). 
The X-ray excess and the slight temperature increase in the region 
between the two clusters indicate a physical link between 
this pair of clusters (e.g. Fujita et al. 1996,
Fabian et al. 1997, Markevitch et al. 1998). 
Recent XMM X-ray analyses (Sakelliou \& Ponman 2004, Bourdin \& Mazzotta 2008)
show that neither of the two clusters 
contains a cooling core. However, the reasonably relaxed morphology of the clusters 
and the absence of major temperature anomalies argue against models in which A399 and A401 
have already experienced a close encounter. The link region between A399 and A401 has been studied with the Suzaku 
satellite by Fujita et al. (2008). The metallicity of the hot intergalactic medium in
this region is found to be comparable to those in the inner regions of the clusters.
The authors conclude that the metal enrichment in
this region is likely caused by strong winds from galaxies
before the cluster formation rather than by cluster collision.
These analyses indicate that the clusters are just
starting to mildly interact and that the sub-features
found in their inner regions are related to the individual merging histories of each cluster 
separately, rather than to the remnant of a previous merger of the two systems.
Indeed, in the light of all these indications, it  
seems likely that A399 and A401 are two merger remnants, just before they merge together to form a 
single rich cluster of galaxies.
 
A401 is already known to contain a faint radio halo (Harris et al. 1980, Roland et al. 1981, Giovannini 
et al. 1999, Bacchi et al. 2003), while a hint for a diffuse emission in A399 has been suggested by 
Sakelliou \& Ponman (2004) based on the NRAO VLA Sky Survey (NVSS) images. In this letter, we report on the results of 
deeper observations performed at 1.4 GHz with Very Large Array (VLA) in C and D configurations.

Throughout this paper we assume a $\Lambda$CDM cosmology with $H_0$ = 71 km s$^{-1}$Mpc$^{-1}$,
$\Omega_m$ = 0.27, and $\Omega_{\Lambda}$ = 0.73. At the distance of A399, 1\arcsec~ corresponds to 1.35 kpc.

\section{Radio observations of A399}\label{sec:a399}

The cluster of galaxies A399 was observed with the VLA at 1.4 GHz in the C and D configurations 
 on April and August 2004, respectively (program AS791). The two observations have the
 same pointing, RA=02$^{h}$57$^{m}$54$^{s}$ and DEC=+13\degr 01\arcmin 34\arcsec (J2000). They also share
 the same observing IF frequencies, 1415/1465 MHz, and bandwidth, 50 MHz.
The total time on source is of 2.2 and 1.5 hours in C and D configurations, respectively.

Calibration and imaging were performed with the NRAO
Astronomical Image Processing System (AIPS).
The data were calibrated in both phase and amplitude.
The phase calibration was completed by using the nearby secondary calibrator 0321+123  
observed at intervals of $\sim$30 minutes. 
The flux-density scale was calibrated by observing 0137+331 (3C\,48). 
Total intensity images were produced
following the standard procedures: Fourier-Transform, 
Clean, and Restore. Several cycles of self-calibration were applied in order to remove residual
phase variations.

From the D configuration data, we produced an image with a circular FWHM beam 
of $45\arcsec\times 45\arcsec$~ and a noise of 40 $\mu$Jy/beam ($1\sigma$ level).
The C configuration data resulted in an image with an FWHM beam 
of $14\arcsec\times 13\arcsec$~ and a noise level of 40 $\mu$Jy/beam.

\section{Results}

\begin{figure*}[t]
\centering
\includegraphics[width=18 cm]{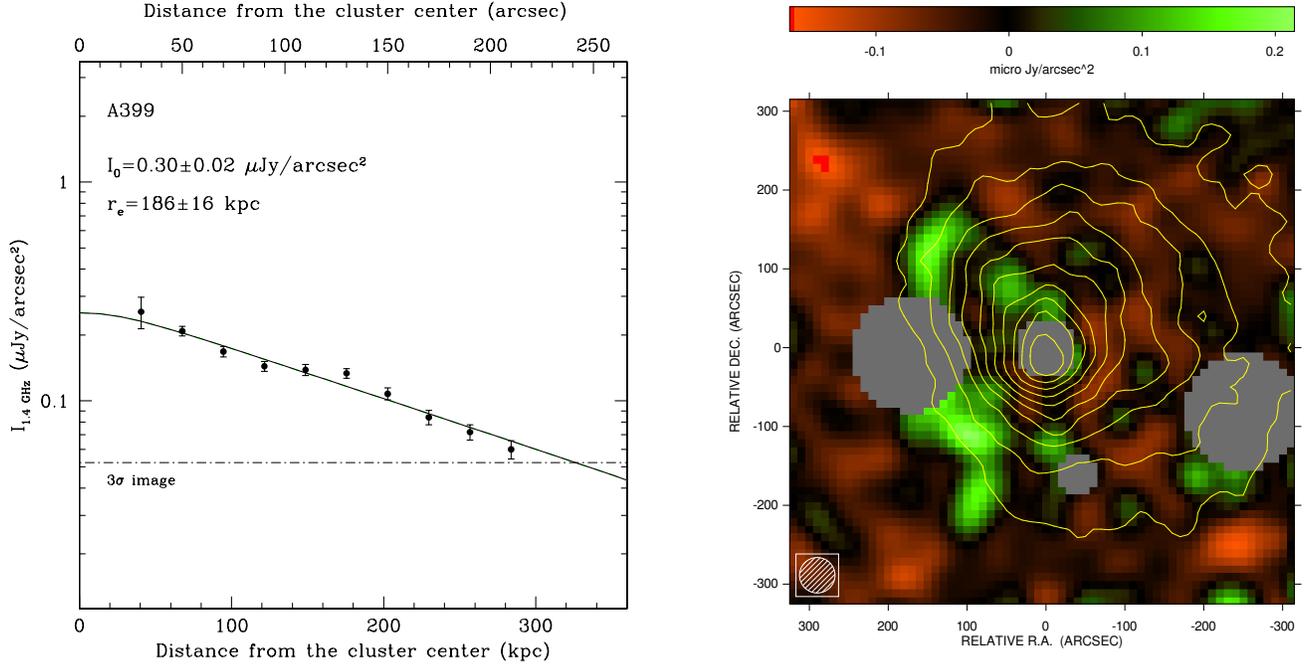}
\caption{
Left: The azimuthally averaged brightness profile of the radio halo emission
in A399. The profile is calculated in concentric annuli and all the discrete sources have been masked.
The horizontal dashed-dotted line indicates the 3$\sigma$ noise level of 
the radio image while the continuous line indicates the best fit profile described by an exponential law
(see text). Right: Overlay of the XMM X-ray contours on the fit residuals, gray regions correspond to the masked
 discrete radio sources.}
\label{fig2}
\end{figure*}

\subsection{A new radio halo in A399}

We present the new radio image of A399, taken with the VLA in D 
configuration at 1.4 GHz, in the left panel of Fig.\ref{fig1}. The 
radio iso-contours are overlaid on the MOS1+MOS2 XMM-Newton image of the 
cluster X-ray emission in the 0.2$-$12 keV band (Obs. id: 0112260101). 
The relatively large beam of the radio observation provides an optimal 
S/N ratio to the low-surface brightness emission. We find that the 
central region of A399 is permeated by a very low-surface brightness 
diffuse emission which is classified as a radio halo. The halo has an 
angular size, as measured from the $3\sigma$ radio isophote, of about
 420\arcsec. The corresponding linear size is $LLS\simeq570$ kpc. The 
radio halo is significantly smaller than the maximum angular size 
visible by the VLA D array at this frequency, i.e. 900\arcsec. This 
ensures that we should have recovered most of the flux density from the 
extended radio emission.  The morphology of the halo is quite 
regular, although an arc-like feature (indicated by a dashed curved line in 
Fig.\ref{fig1}) of enhanced radio emission is observed in 
correspondence of the sharp X-ray edge to the east of the cluster core 
reported by Sakelliou \& Ponman (2004) and Bourdin \& Mazzotta (2008). 
The X-ray surface brightness left to the edge cut offs sharply and, as a 
consequence, the overall X-ray cluster morphology appears to be elongated 
to the north-west. Sakelliou \& Ponman (2004), based on the results of the 
numerical simulations from Takizawa (1999), attributed the origin of the 
X-ray edge to the infall of a low mass system travelling east to west into 
A399. The radio feature could be originated by a relatively faint shock in the
intracluster medium due to the merging of this low mass system.

We used the higher resolution of the C configuration image to separate 
the diffuse cluster emission from that of the unrelated discrete radio 
sources. A zoom of the inner $8\arcmin \times8\arcmin$ region of A399 is 
shown in the right panel of Fig.\,\ref{fig1}. The total intensity radio 
contours of the VLA observation in C array are overlaid on the DSS2 red 
plate image\footnote{See http://archive.eso.org/dss/dss}
 of the cluster. The C configuration image indicates clearly 
that the radio halo is not due to the blending of discrete sources. In 
fact, we can identify just four discrete radio sources seen in 
projection over the halo. The source on the left of the halo (labeled A 
in Fig.\,\ref{fig1}) has a flux density of $S_{1.4}\simeq 2.5\pm0.07$ 
mJy. There is no optical counterpart for radio source A in the DSS2 
image. A second faint source (labeled B; $S_{1.4}\simeq 0.7\pm0.04$ 
mJy) is visible near the cluster center. The overlay with the optical 
image shows that this discrete radio source is associated to a faint 
galaxy located 14\arcsec~to the north-east
 of the cD galaxy which, on the contrary, appears to be radio quite. 
The small galaxy associated with radio source B has a redshift of 
$z=0.07691$ (Hill \& Oegerle 1993), and therefore it is likely a member 
of A399. A third very faint source (labeled C; $S_{1.4}\simeq 
0.3\pm0.04$ mJy) is located in the south region of the halo. Source 
C has a clear optical counterpart in the DSS2 image but no redshift 
information is available in the literature. Finally, the stronger 
discrete source (labeled D) is located on the extreme right side of the 
radio halo and has a flux density of $S_{1.4}\simeq 36.8\pm 1.1$ mJy. 
There is no optical counterpart for radio source D in the DSS2 
image.

In the left panel of Fig.\ref{fig2} we show the azimuthally averaged 
radio halo brightness profile obtained from the D configuration image at 
45\arcsec~resolution after the correction for the primary beam 
attenuation. Each data point represents the average brightness in 
concentric annuli of half beam width. Discrete sources have been masked 
out and excluded from the statistics. The observed brightness profile is 
traced down to a level of 3$\sigma$. Following Murgia et al. (2009), we 
modeled the radio halo brightness profile, $I(r)$, with an exponential 
of the form $I(r)=I_{0}e^{-r/r_e}$, whose best fit is represented by the 
solid line in Fig.\ref{fig2}. The fit is performed in the image plane as 
described in Murgia et al. (2009). In order to properly take into 
account the resolution, the exponential model is first calculated in a 
2-dimensional image, with the same pixel size and field of view as that 
observed, and then convolved with the same beam by means of a Fast 
Fourier Transform. The resulting image is masked exactly in the same 
regions as for the observations. Finally, the model is azimuthally 
averaged with the same set of annuli used to obtain the observed radial 
profile. All these functions are performed at each step during the fit 
procedure.  As a result, the values of the central brightness, $I_0$, 
and the e-folding radius $r_{e}$ provided by the fit are deconvolved 
quantities and their estimate includes all the uncertainties related to 
the masked regions and to the sampling of the radial profile in annuli 
of finite width. The best fit of the exponential model yields a central 
brightness of $I_0=0.3\pm 0.02$ $\mu$Jy/arcsec$^2$ and $r_{e}=186 \pm 
16$ kpc with a reduced $\chi^2=1.6$. The halo flux density obtained by 
integrating the radio brightness profile up to the $3\sigma$ isophote 
and excluding the discrete sources is of $S_{1.4}=16.0 \pm 2$ mJy, which 
corresponds to a radio power of $P_{1.4}\simeq 2\times 10^{23}$ W/Hz. 
In the right panel of Fig.\ref{fig2}, we show the overlay of the 
XMM X-ray contours on the residuals obtained
by subtracting the best fit from the observed radio halo image. The 
most striking feature emerging in the residuals is the arc-like 
structure mentioned above. The radio arc appears to be coincident, at 
least in projection, with the sharp edge of the X-ray cluster emission.

\begin{figure}[t]
\centering
\includegraphics[width=9 cm]{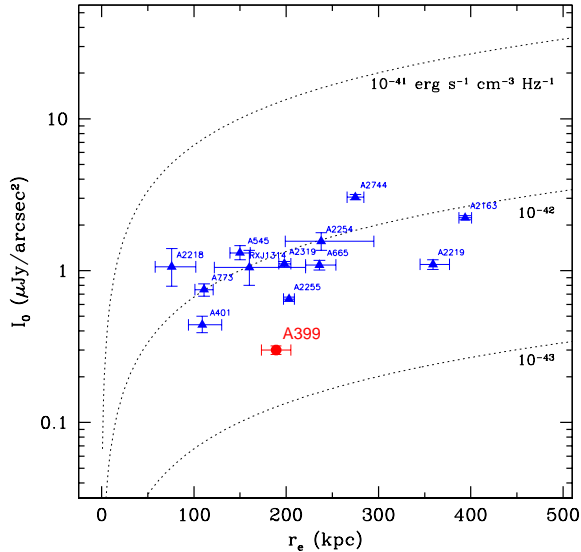}
\caption{ Best fit central brightness $I_0$ versus
the length-scale $r_e$ of A399 in comparison with the set of radio halos analyzed in Murgia et al. (2009). 
The dotted lines indicate regions of constant synchrotron emissivity.
}
\label{fig3}
\end{figure}

In Fig.\ref{fig3}, we show the best fit central brightness $I_0$ versus 
the length-scale $r_e$ of A399 in comparison with the set of radio halos 
analyzed in Murgia et al. (2009). A399 is somewhat fainter than A401 but 
twice as large. These characteristics made it the cluster with the 
lowest synchrotron emissivity $J_{1.4}=1.6\times 10^{-43}$ 
erg\,s$^{-1}$cm$^{-3}$Hz$^{-1}$. Despite its low emissivity, the 
physical properties of the small radio halo in A399 are in good 
agreement with the extrapolation of the properties of the giant 
($LLS$$>$1 Mpc) radio halos presented in Giovannini et al. (2009). In 
particular, the correlation between the LLS and the radio power and 
between the X-ray luminosity and radio power are in good agreement, 
confirming that small size and giant radio halos may have similar origin 
and properties.

\subsection{The double radio halo in the A399 - A401 system}

By combining the new D configuration image of A399 presented in this 
work and the radio image of A401 obtained by Bacchi et al. (2003), we 
produced a wide field image of the A399 - A401 complex. The individual 
images of the two galaxy clusters have been taken with the same VLA 
configuration and frequency. They are characterized by the same beam, 
cellsize, and a very similar noise level. Moreover, the two galaxy 
clusters are located at a projected distance of 36$\arcmin$. 
Incidentally, this is exactly the size of the primary beam of the
 VLA antennas at 1.4 GHz. An accurate mosaicing would require the 
pointings to be spaced by half the primary beam size, however, for 
illustrative purposes, we combined the two images in a single field that 
comprises both galaxy clusters through the AIPS task LTESS.

Similarly, we produced a mosaicing of the X-ray emission in the 
0.2$-$12 keV band by combining three different XMM pointings centered on 
A399 (Obs. id. 0112260101), A401 (Obs. id. 0112260203), and in between 
the two clusters (Obs. id. 0112260302). We used the tool reproject\_image\_grid, 
implemented in the CIAO software (Fruscione et al. 2006), to reproject 
the MOS1 and MOS2 images of the three pointings to a common frame of 
reference, creating a mosaic from the individual observations. The final 
X-ray image has been exposure corrected and smoothed with a Gaussian 
kernel of $\sigma=12\arcsec$.

\begin{figure*}[t]
\centering
\includegraphics[width=16 cm]{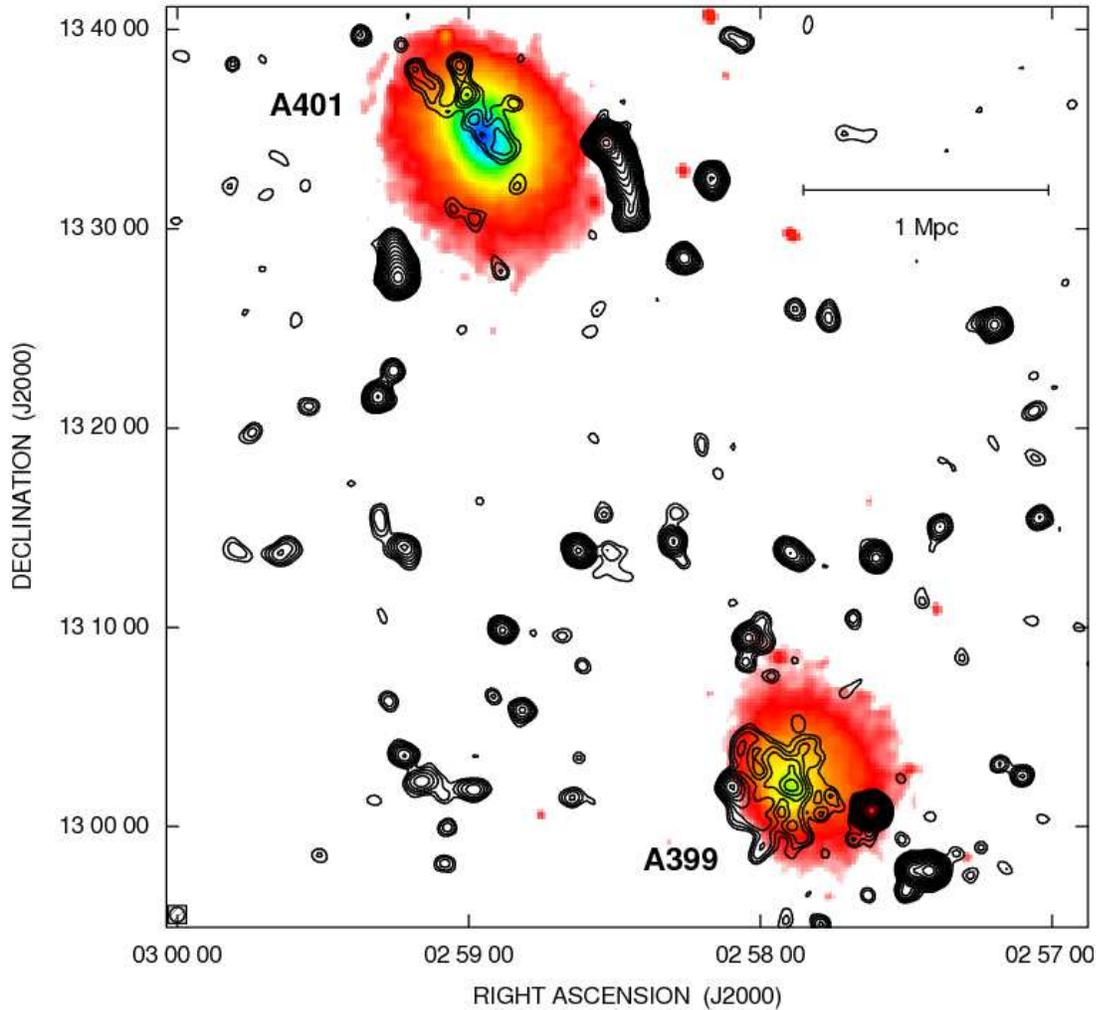}
\caption{
Total intensity radio contours of the complex A401-A399,
obtained by combining the two VLA observations at 1.4 GHz in D configuration of the two clusters, as described in the text.
The radio image is shown by the iso-contours and has an FWHM of 45$''\times 45''$.
The first contour level is drawn at 120 $\mu$Jy/beam and the rest are spaced by a factor of $\sqrt{2}$.
The sensitivity (1$\sigma$) is 40 $\mu$Jy/beam.
Total intensity radio contours are overlaid on the XMM X-ray image in the 0.2$-$12 keV band. 
The X-ray image has been convolved with a Gaussian kernel of $\sigma=12\arcsec$. 
}
\label{fig4}
\end{figure*}

In Fig.\,\ref{fig4}, we present the iso-contours of the radio mosaic 
overlaid on the mosaic of the X-ray images. The figure clearly shows the 
diffuse emission in the radio and X-ray bands originating from the 
intracluster medium of the two clusters. Indeed, we can conclude that 
the close pair of galaxy clusters A401 and A399 can be considered as the 
first example of double radio halo system.

The small radio halos in A339 and A401 are among the faintest so far 
seen in cluster of galaxies (see also Fig.\,\ref{fig3}). In A399,  
the centroid of the radio halo is shifted with respect to the X-ray peak  
 while the halo in A401 has an elongated shape in the same direction of the X-ray emission. 
Although in a number of clusters a spatial correlation is observed between the radio 
halo brightness and the X-ray emission (Govoni et al. 2001), the 
distortions and offsets of the radio emission in A399 and A401 seem to 
be common properties of small halos. Feretti et al. (2009) analyzed the 
positions of radio halos with respect to the X-ray gas 
distribution. Both giant and small radio halos can be significantly
 shifted, up to hundreds of kpc, with respect to the centroid of the 
host cluster. Moreover, they found that this effect becomes more 
relevant when halos of smaller size, like those in A399 and A401, are considered.

\section{Discussion}

We briefly discuss the discovery of the double radio halo in the A401 - 
A399 system in relation to the evolutionary state of the merger between 
the two clusters.

\subsection{Post-merger scenario} 

We first consider the scenario in which A399 and A401 are in a 
post-merger state as early suggested by Fabian et al. (1997). In this 
case, both radio halos may have formed as a consequence of the 
dissipation of the energy delivered in a close encounter between the two 
clusters. In this context, it is interesting to compare the life-time of 
the synchrotron emitting cosmic ray electrons, $t_{syn}$, 
with the time elapsed since the clusters ``collision'', $t_{coll}$.
The radiative losses due to the inverse Compton scattering
of the cosmic microwave background photons limit the age synchrotron
electrons radiating at 1.4\,GHz to $t_{syn}\lesssim 10^8$ yrs. 
According to Yuan et al. (2005), the encounter event occurred 
$t_{coll}\sim 3.5 \times 10^{9}$ yrs ago and the two clusters are 
currently separating apart at a true relative velocity of about 580 
km/s. Hence,  $t_{coll}\gg t_{syn}$. This implies that the continuing 
turbulent dissipation of the energy released during the previous 
encounter between the two clusters should be able to maintain the two 
radio halos visible long after the 
collision. However, this amount of time seems to be much longer 
than the typical time-scale for the dissipation of turbulence 
in the intracluster medium during cluster mergers, which is $\sim 1$ Gyr 
(Brunetti et al. 2009). Therefore, the post-merger scenario seems not 
applicable to explain the origin of the double radio halo.

\subsection{Pre-merger scenario}
The recent X-ray analyses of the A399 and A401 
complex by Sakelliou \& Ponman (2004) and Bourdin \& Mazzotta (2008) 
disfavour the above scenario and rather indicate that the clusters are 
just at an early stage of merging and that they are interacting mildly, 
because their separation is too large for more dramatic effects. In this 
scenario, the origin of the double radio halo has to be attributed to 
the individual merging histories of each cluster separately, rather than 
to the result of a close encounter between the two systems. The two 
clusters must have experienced a similar sequence of mergers which prevented the
 formation of the cool cores and left significant 
residual kinetic energy in the gas to power the radio halos. 
If the two clusters have formed within a collapsing filament, and are now falling 
towards one another, then this could increase the probability of the 
observed association considerably, since they are no longer independent 
systems. 
The origin of the double radio 
halo in A399 and A401 would be indeed a direct consequence of the same environment 
in which the two clusters formed. This has important implications on the models
on the origin of small radio halos since it implies that minor merger  
play a fundamental role in the formation and maintaining of these diffuse radio sources. 

It interesting to note that neither A399 nor A401 are cool core clusters.
According to the results of the numerical simulations by Burns et al. (2008), many noncool 
core clusters undergo major merger early in their history. Smaller
mass units with cool cores continue to be accreted by these systems but are 
usually disrupted by ram pressure within a single core passage. 
This is what we are probably witnessing in the case of A399. The sharp X-ray edge
observed in the X-ray image has been interpreted by Sakelliou \& Ponman (2004)
as the result of the infall of a smaller sub-unit. In this work we showed 
that this feature appears to be strictly connected with a bright arc-like structure in
the periphery of the radio halo.
There are also indirect evidence that A401 is disturbed by a merger since
it is elongated, hosts a hot core, and has a complex temperature distribution. 

\subsection{Tidal interaction}
Finally, we may suggest a third scenario 
in which the formation of the double radio halo is triggered by the long-distance tidal 
interactions occurring between the two clusters. The presence of non-thermal radio emission could 
prove that, even if the two sub-clusters are still two distinct 
entities, their mutual interaction happens to be already experienced by 
their intracluster media, and would therefore indicate that merger 
effects can be at work at an early stage of the process. However, the tidal 
forces should be relevant only in the external regions of the clusters, and not 
 in the central regions occupied by the radio halos, as already noted by Sakelliou \& Ponman (2004). 
Furthermore, it is not clear what mechanism would be able to turn a small tidal acceleration into 
relativistic particles. Hence, it seems unlikely that the tidal scenario could provide an
acceptable explanation for the double radio halo in A401 $-$ A399.

\section{Conclusions}

In summary, we find that the central region of A399 is permeated by a 
diffuse low-surface brightness radio emission which is classified as a 
radio halo with a length-scale of about 190 kpc and a central brightness 
of 0.3 $\mu$Jy/arcsec$^2$. Indeed, given their projected distance of 
$\sim 3$ Mpc, the pair of galaxy clusters A401 and A399 can be 
considered as the first example of double radio halo system.

The origin of the double radio halo is likely attributed to 
the individual merging histories of each cluster separately, rather than 
to the result of a close encounter between the two systems. 
The two clusters must have experienced a similar sequence of mergers 
which prevented the formation of the cool cores and left significant 
residual kinetic energy in the gas to power the radio halos.

It seems likely that the two small radio halos in A399 and A401 are in proximity 
to merge together eventually forming a single giant radio halo or a 
large scale diffuse emission identified with a filamentary 
supercluster-like structure. This offers new clues to how giant radio halos could form, since
it is now clear that magnetic fields and relativistic particles can already be present in the
intracluster medium of two clusters well {\it before} a major merger event between them.

\begin{acknowledgements}
We thank the referee for providing useful comments and suggestions.
We are also grateful to Trevor Ponman for his valuable comments on the original draft and to Andrea Possenti for 
very useful discussions. This research was partially supported by ASI-INAF I/088/06/0 -
High Energy Astrophysics and PRIN-INAF 2008.
The National Radio Astronomy Observatory (NRAO) is a facility of the National Science Foundation, 
operated under cooperative agreement by Associated Universities, Inc. This work is based on observations obtained with XMM-Newton, 
an ESA science mission with instruments and contributions directly funded by ESA Member States and the USA (NASA). 
\end{acknowledgements}


\begin{thebibliography}{}

\bibitem{}
Bacchi, M., Feretti, L., Giovannini, G., Govoni, F.\ 2003, A\&A, 400, 465

\bibitem{} 
Bourdin, H., \& Mazzotta, P.\ 2008, \aap, 479, 307 

\bibitem[]{} Brunetti, G., Cassano, R., Dolag, K., \& Setti, G.\ 2009, arXiv:0909.2343 

\bibitem{}
Buote, D.~A.\ 2001, \apjl, 553, L15

\bibitem{}
Burns, J.~O., Hallman, E.~J., Gantner, B., et al.\ 2008, \apj, 675, 1125

\bibitem{} 
Fabian, A.C., Peres, C.B., White, D.A., 1997, \mnras, 285, L35 

\bibitem{} 
Feretti, L., Bonafede,  A., Giovannini, G., Govoni, F., \& Murgia, M.\ 2009, arXiv:0910.1519 

\bibitem{} 
Fujita, Y., Koyama, K., Tsuru, T., \& Matsumoto, H.\ 1996, \pasj, 48, 191 

\bibitem[]{} 
Fujita, Y., Tawa, N., Hayashida, K., Takizawa, M., Matsumoto, H., Okabe, N., \& Reiprich, T.~H.\ 2008, \pasj, 60, 343 

\bibitem[]{} 
Fruscione, A., et al.\ 2006, \procspie, 6270

\bibitem{}
Giovannini, G., Tordi, M., \& Feretti, L.\ 1999, New Astronomy, 4, 141

\bibitem{} 
Giovannini, G., Bonafede, A., Feretti, L., et al., A\&A in press, arXiv:0909.0911 

\bibitem{}
Govoni, F., En{\ss}lin, T.~A., Feretti, L., \& Giovannini, G.\ 2001, \aap, 369, 441 

\bibitem{} 
Harris, D.E., Kapahi, V.K., Ekers, R.D., 1980, \aaps, 39, 215 

\bibitem{} 
Markevitch, M., Forman, W.R., Sarazin, C.L., Vikhlinin, A., 1998, \apj, 503, 77 

\bibitem{} Murgia, M., Govoni, F., Markevitch, M., et al. \ 2009, \aap, 499, 679 

\bibitem{} 
Oegerle, W.~R., \& Hill, J.~M.\ 2001, \aj, 122, 2858 

\bibitem{}
Reiprich, T.~H., B{\"o}hringer, H.\ 2002, \apj, 567, 716

\bibitem{} 
Roland, J., Sol, H., Pauliny-Toth, I., Witzel, A., 1981, \aap, 100, 7 

\bibitem{} 
Sakelliou, I., Ponman, T.J., 2004, \mnras, 351, 1439 

\bibitem[]{} Takizawa, M.\ 1999, \apj, 520, 514 

\bibitem[]{} 
Yuan, Q.-R., Yan, P.-F., Yang, Y.-B., \& Zhou, X.\ 2005, Chinese Journal of Astronomy and Astrophysics, 5, 126 

\end{thebibliography}
\end{document}